# Effects of thermal treatment on radiative properties of HVPE grown InP layers

Serge Luryi[1], Oleg Semyonov, Arsen Subashiev,
*ECE Department, State University of New York, Stony Brook, NY 11794-2350 USA*

Joseph Abeles, Winston Chan, Zane Shellenbarger,
*SRI International, 201 Washington Rd., Princeton, NJ 08540 USA*

Wondwosen Metaferia and Sebastian Lourdudoss,
*KTH-Royal Institute of Technology, 16440 Kista, Sweden*

**Abstract.** Radiative efficiency of highly luminescent bulk InP wafers severely degrades upon heat treatment involved in epitaxial growth of quaternary layers and fabrication of photodiodes on the surface. This unfortunate property impedes the use of bulk InP as scintillator material. On the other hand, it is known that thin epitaxial InP layers, grown by molecular beam epitaxy (MBE) or metal–organic chemical vapor deposition (MOCVD), do not exhibit any degradation. These layers, however, are too thin to be useful in scintillators. The capability of hydride vapor phase epitaxy (HVPE) process to grow thick bulk-like layers in reasonable time is well known, but the radiative properties of HVPE InP layers are not known. We have studied radiative properties of 21 µm thick InP layers grown by HVPE and found them comparable to those of best luminescent bulk InP virgin wafers. In contrast to the bulk wafers, the radiative efficiency of HVPE layers does not degrade upon heat treatment. This opens up the possibility of implementing free-standing epitaxial InP scintillator structures endowed with surface photodiodes for registration of the scintillation.

## 1. Introduction

Among A3B5 semiconductor materials Indium Phosphide has remarkably versatile properties that makes it useful for a number of applications including nanowire array solar cells [1,2], semiconductor scintillator for gamma-detection [3-5], detectors of neutrino [6], and, ultimately, the whole range of photonic integrated circuits [7,8], comprising optical waveguides, amplifiers, phase modulators, etc. This ultra-wide range of applications is predominantly due to the possibility of growing on InP different lattice-matched quaternary material layers with properly tuned bandgaps

---

[1] Corresponding author, Serge.Luryi@stonybrook.edu, Tel. (631) 632-8420, Fax (631) 632-8494

complemented by well-developed etching technologies. Optoelectronics applications can use advanced epitaxial growth technologies, such as molecular beam epitaxy (MBE) or metal–organic chemical vapor deposition (MOCVD). However, scintillator applications require large-volume crystals and this requirement renders MBE and MOCVD less suitable.

Appropriate quality bulk wafers are commercially available [9,10]. Moreover, our studies of optical and luminescent properties of InP Acrotec wafers [9] showed uniquely high radiative efficiency exceeding 98% at room temperature for moderately *n*-doped crystals [11,4]. Further studies, however, revealed a fatal degradation of the high-efficiency wafers after high-temperature treatment, unavoidable in subsequent epitaxial growth of lattice-matched photodiodes.

The degradation of high-radiative efficiency InP wafers [9] was first observed when the wafers were endowed with surface photodiodes. The process [12] involves two high-temperature steps: the epitaxial growth of quaternary InGaAsP ($E_G$ = 1.24 eV) layers at 600C and the Zn diffusion at 525C. The photoluminescence intensity in the processed wafers was down by nearly two orders of magnitude when compared with that in virgin wafers. The exact mechanism of the degradation is not established but presumably is due to deep centers that show themselves in the Arrhenius plot of the luminescence intensity degradation with the temperature of treatment.

Such centers apparently do not exist or do not manifest themselves in *epitaxial layers* of InP that work so well in optoelectronic applications. Similarly, they do not show up in epitaxial GaAs layers that feature the highest radiative efficiency ever reported in semiconductors [13]. Whatever may be the cause of the thermal degradation of luminescence, it is not expected to operate in epitaxial material. It is known that the luminescent properties of low-doped epitaxial layers, as opposed to those of a bulk Czochralski-grown wafer, do not degrade under high-temperature treatment.

Unfortunately, the widely available epitaxial techniques like MBE and MOCVD cannot be used to produce really thick free standing structures. At a "fast" growth rate of 1 µm an hour it would take six weeks of continuing growth to produce a millimeter-thick structure. The direct road to the implementation of epitaxial InP scintillator is to use the hydride vapor phase epitaxy (HVPE) technology that can afford growth rates exceeding 100 µm an hour while retaining high-quality of thick grown layers [14]. HVPE grown "quasi-bulk" material can be expected to have non-degrading luminescent properties.

Here we report the HVPE growth of InP layers with thickness of 21 μm with subsequent studies of its luminescent properties before and after a generic high temperature treatment (represented by MOCVD of an additional thin InP layer). We found no tangible degradation of efficiency.

## 2. Luminescence degradation of InP bulk wafers upon heating

We have investigated the changes in luminescence upon wafer anneals at different temperatures and different environments. Some of the anneal experiments were carried out in MOCVD reactor at SRI International. Figure 1 shows the typical photoluminescence spectra "before" and "after" the annealing cycle. Our typical exposures lasted about half an hour, but the observed degradation was not sensitive to the *time* of the wafer exposure to high temperatures. The degradation was fully in place even after 10 min exposures, as was confirmed by random checks.

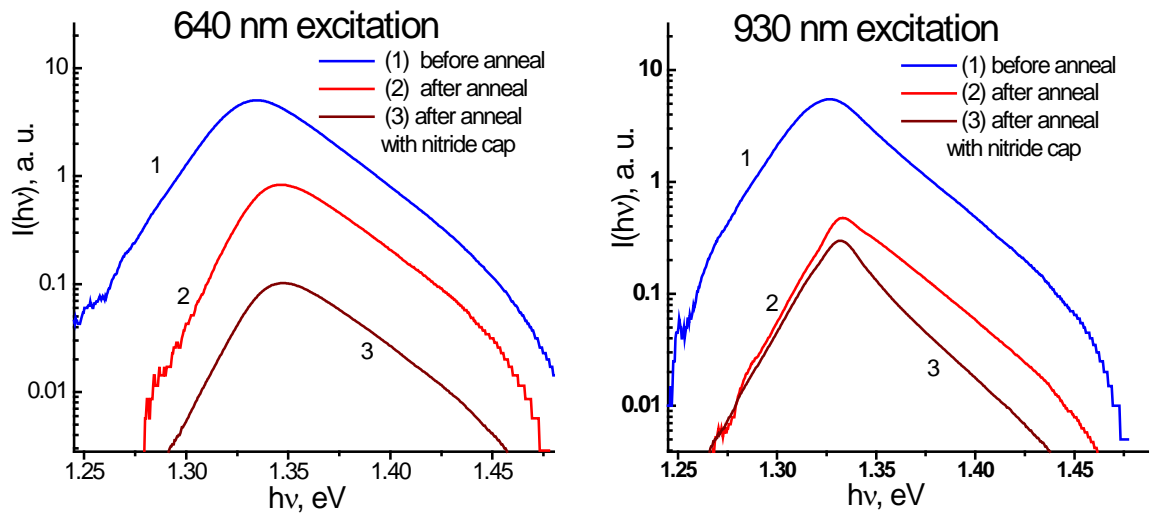

FIG. 1. Photoluminescence in moderately-doped (S: $3 \times 10^{17}$ cm$^{-3}$) *n*-InP samples before and after high-temperature anneal at 550C for two excitation energies. Luminescence in the annealed samples is strongly degraded, both when excited with red light (excitation near the surface with penetration depth ~ 200 nm) and with penetrating radiation of wavelength 930 nm (penetration depth ~ 10 μm). In this and all other presently reported experiments, the intensity of excitation was kept at the level corresponding to low injection [16].

Note the blue-side shift of the luminescence peak after the anneal by 12 meV relative to that before anneal. We attribute this shift to smaller hole diffusion length and therefore smaller escape length for the luminescent radiation. Top surface and backside excitation geometries give similar results.

The use of nitride coating apparently does not protect the material from degradation. In fact, luminescence in nitride-coated samples shows even higher degradation, with similar peak shift. Evidently, either some impurities helpful to high radiative efficiency, e.g. hydrogen (known for defect passivation in InP) disappear from the sample, or some new defects emerge, quenching the lifetime. The degradation is definitely not a surface effect, as is evident from photoluminescence experiments with the highly penetrating 930-nm radiation that excites carriers far from the surface (penetration depth ~ 10 μm). The lifetime measurements [11] have also confirmed that the degradation is not a surface effect.

Temperature dependence of the degradation factor $D(T)$ (defined as the ratio of luminescence in the virgin wafer to that after anneal at temperature $T$) is shown by the Arrhenius plot in Fig. 2. We have found that the degree of degradation is exponential, $D(T) \sim \exp(-E_A/kT)$ with the activation energy $E_A = 1.5$ eV in the high-temperature region ($T > 500C$). This activation energy is consistent with the hypothesis of H out-diffusion and loss through the surface.

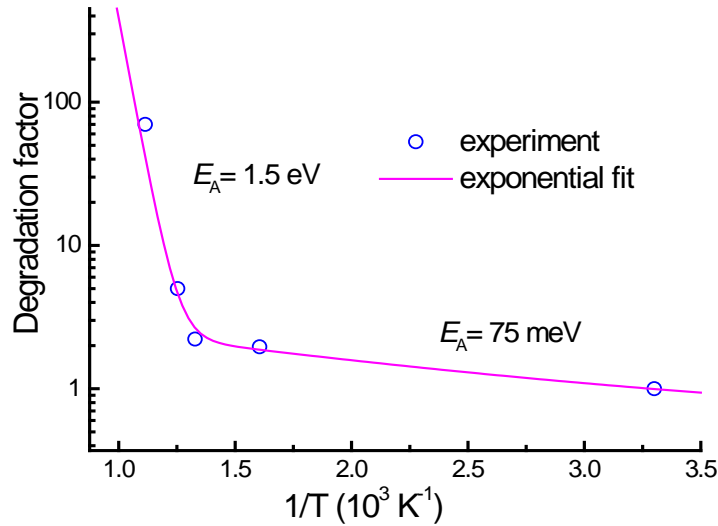

FIG. 2. Temperature dependence of luminescence degradation in Acrotec InP samples. The activation energy $E_A$ is defined by $D(T) \sim \exp(-E_A/kT)$ and is evidently different at low anneal temperatures ($T < 500C$) and high-temperature region ($T > 500C$).

Apparently, irreversible degradation happens near 500C. The gentle exponent below 500C is associated with surface degradation of InP wafer, whereas the steep exponent above 500C could be due to depletion of H in the interior.

An indirect support to the latter hypothesis is provided by the lateral distribution of the photoluminescence intensity in a virgin wafer. Before the wafer was cut it was part of a long cylindrical ingot that was subject to very high temperatures (>950C). As the cylinder cools down, it should become depleted of hydrogen near its surface. Therefore, one could expect luminescence degradation in a small annulus near the wafer edge that was part of the cylindrical surface. We looked for this effect and found it: within one millimeter from the edge of the virgin wafer, the photoluminescence intensity goes down, almost by an order of the magnitude.

We have made no attempt to quantify the loss of hydrogen with direct measurements by SIMS and the hydrogen hypothesis remains a hypothesis. The exact mechanism of degradation has not been established in this work. Attempts to remedy the degradation (such as hydrogen passivation with the hope of "rejuvenating" the luminescence, coating the samples during the high-temperature anneal with silicon nitride, as well as annealing the sample for 30 minutes in the MOCVD reactor in a hydrogen-rich atmosphere) were unsuccessful.

## 3. Temperature tolerance of HVPE-grown InP layers

The HVPE layers were grown in the Laboratory of Semiconductor Materials at KTH-Royal Institute of Technology, Sweden. For a preliminary experiment the approximately 21 μm thick  layers of moderately n-doped InP (sulfur-doped with $3 \times 10^{17}$ cm$^{-3}$ concentration, similar to our best bulk Acrotec wafer) were grown on a heavily doped InP substrate. The growth was done at 610C with V/III ratio of 10 (i.e., [PH3]/[InCl]=10) for 2.5 hrs. Morphology of the growth was not ideal with visible surface defects. The visible defects are most likely due to the particles from extraneous wall deposition caused by a relatively long growth time. We have not observed such defects when growth was done for shorter time (up to 1.5 hrs). Thus, the morphological quality of the layer can be improved and thicker layers can also be grown by optimizing the growth conditions.

The results for the luminescence spectra recorded in the reflection geometry for this sample with 850 nm excitation are presented in Fig. 3, together with spectra of the heavily-doped ($n \sim 4\text{-}6 \times 10^{18}$ cm$^{-3}$) substrate and the bulk virgin Acrotec InP wafer (S-doped at $3\times10^{17}$ cm$^{-3}$).

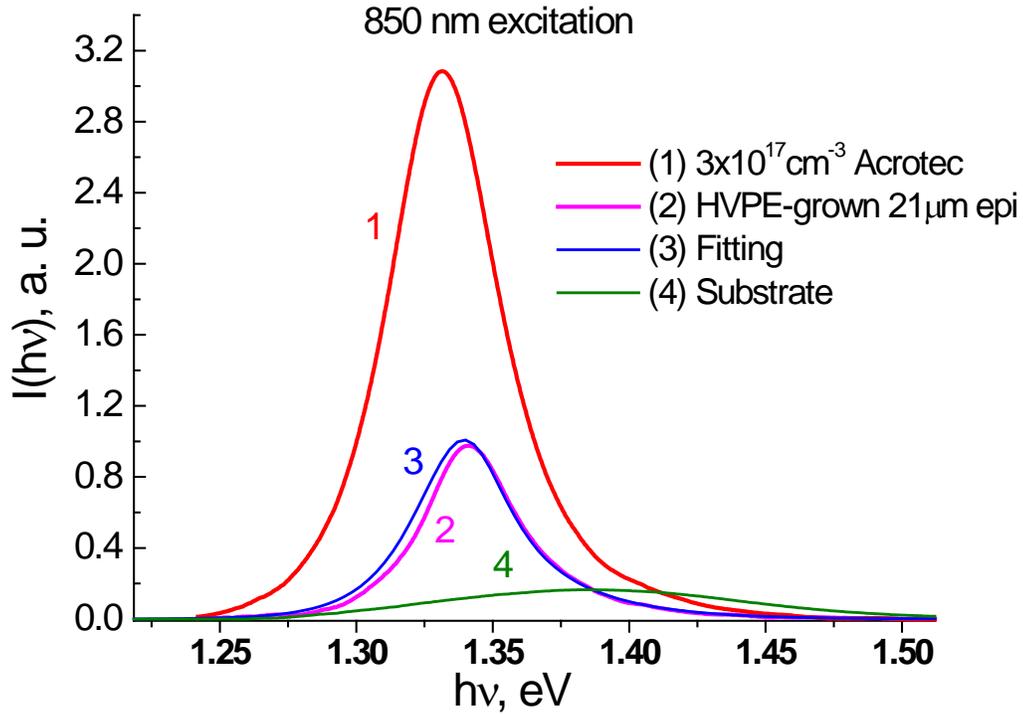

FIG. 3. Luminescence spectra of HVPE-grown 21 μm thick layer (S-doped at $3\times10^{17}$ cm$^{-3}$) recorded in reflection geometry and compared to the spectra of similarly-doped virgin Acrotec wafer and to theoretical fitting that accounts for quenching of luminescence by the heavily-doped substrate ($n \sim 4\text{-}6 \times 10^{18}$ cm$^{-3}$). Also shown are the substrate spectra measured separately.

Most notable is the high luminescence intensity that is only 4 times smaller than that of the best virgin bulk (Acrotec) wafer. The blue shift of the line shape and some enhancement at the blue wing of the line are also very close to what can be expected for the layer of this thickness. Thus, the luminescent properties of the HVPE layer are similar to our best bulk wafers of similar doping. Despite the 610ºC epitaxial growth temperature, there has been no degradation of luminescence.

For the ultimate check of the temperature endurance of the epitaxial layer we did complementary experiments with epitaxial samples, supplying them with an additionally grown undoped or moderately doped epitaxial layer of thickness $t = 0.2$ µm. The purpose of these experiments was to ascertain that we can use HVPE material as a substrate for subsequent MOCVD of surface detector structures – hopefully with no further degradation of luminescence in HVPE layers.

To this end, the HVPE wafer – comprising 21-µm $n$-InP epitaxial layer ($n = 3 \times 10^{17}$ cm$^{-3}$) on $n$-InP substrate ($n = 4$ to $6 \times 10^{18}$ cm$^{-3}$) – was cut into quarters. One quarter sample was processed and passed through the conventional procedure of epitaxial growth of an additional 0.2 µm undoped InP layer at a temperature of 650 °C. After the growth, the luminescence radiation of the processed sample (in reflection geometry) was spectrally scanned and compared with the reflection luminescence spectrum of one of the intact "as HVPE grown" samples.

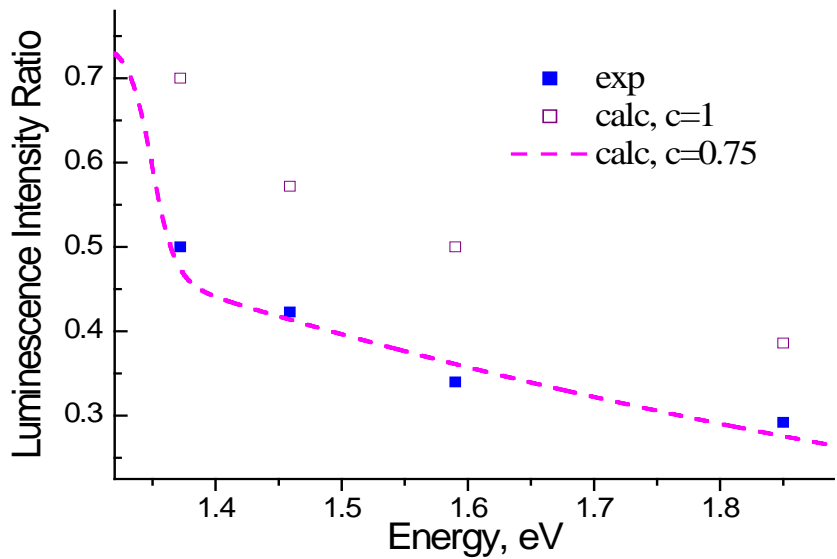

FIG. 4. Ratio of the luminescence intensities of spectra of a HVPE-grown layer with overlayer grown at high temperature relative to those of the virgin HVPE-grown layer as a function of the excitation energy. Open squares show the calculated ratio taking into account only absorption of the excitation light in the overlayer, dotted line shows results of calculations allowing rapid recombination of holes in the overlayer.

Four lasers of different wavelengths (namely 670 nm, 780 nm, 850 nm, and 904 nm) were used for excitation. The excitation laser beam was focused on the surface of the samples on sides with the grown epitaxial layers. A portion of luminescence radiation emerged from the samples was captured by a lens installed at an angle of 30° to the normal of sample's surface and delivered by the optical fiber guide to a scanning monochromator Oriel Cornerstone 260 equipped with the silicon sensor EO Series 5T. The ratios of luminescence integral intensity from the processed and 'as grown' samples calculated for every excitation wavelength are shown in Fig. 4 (after correction of the recorded spectra for detector sensitivity and spectral transmission of the optical system).

## 4. Discussion of the results

Firstly, we stress that the reduction of the luminescence intensity of the epitaxial layer (as compared to the moderately doped thick wafer) does not mean a smaller radiative efficiency – since in a homogeneous wafer the observed luminescence comes from a much thicker region than $d = 21$ µm. In fact, the effective (photon enhanced) diffusion length for the homogeneous wafer [15] was close to $l_{eff} = 80$ µm, which gives a rough estimate for the ratio. This large value of $l_{eff}$ due to photon recycling enabled us to stay within low-injection condition at relatively high excitation power [16].

The active layer contribution is conclusively established by analyzing the position of the luminescence maximum, which shifts as a result of the filtering by reabsorption and is determined by the relation $\alpha(E)d \cong 1$ (see our recent paper [17] for the details). For a thin layer there is less filtering and the maximum shifts to higher energy. Contribution of the highly-doped substrate luminescence is at shorter wavelengths and it does not contribute to the observed spectrum. Therefore, a minor reduction of the luminescence intensity (less than 3-fold for all excitation wavelengths) indicates negligible thermal degradation of the HVPE-grown layer, since the degradation effects, if they exist, are much stronger. The observed reduction of the luminescence for the samples with overgrown layer of thickness $t = 0.2$ µm is apparently due to the fact that the additional layer may have much smaller radiative efficiency and acts as a "dead layer", the region where the holes do not recombine with electrons radiatively.

The fact that overgrown layer has low radiative efficiency is as expected for an undoped layer, where the radiative recombination is suppressed by the absence of majority carriers. For a moderately $n$-doped overlayer this suppression may come as a result of high concentration of surface defects that prevented the growth of thicker HVPE epitaxial layers. With further optimization of the HVPE growth conditions, we do not expect any problem in growing quality MOCVD layers on HVPE substrate.

The effect of the nonradiative layer on the luminescence intensity depends on the absorption of the excitation light. This can be taken into account by an attenuation factor $f_a = \exp[-\alpha \, (E_{ex}) \, t]$. Calculated values of $f_a$ are shown in Fig. 4. With the well-known [18] dependence $\alpha \, (E_{ex})$, the factor accounts for the decline of the ratio at high energies.

Still closer fitting of the experimental data is obtained using the luminescence attenuation factor $f$ in the form $f(E_{ex}) = c \, f_a$ with an additional factor $c$ of about 0.7, common for all 4 excitation energies (see Fig. 4). The origin of the factor can be interpreted as follows. First of all, we note that for all 4 excitation energies, the hole distribution remains the same, since the observed spread of the holes in moderately doped InP wafers is much larger than the absorption length (inverse absorption coefficient at the excitation energy), that remains in a micron range even for the smallest excitation energy, 1.37 eV. Then, the factor $c$ could not be explained by reabsorption of outgoing radiation in the overgrown layer because it would result in a spectral filtering of outgoing radiation leading to a redshift of the line compared to an original luminescence spectrum in reflection geometry. This filtering effect should be clearly noticed in the blue wings. The absence of the redshift suggests that the emission comes predominantly from the regions equally remote from the surface both in case of original and overgrown samples. However, due to the nonradiative layer the effective surface recombination rate for the holes is enhanced, compared to the original geometry in which it is somewhat reduced due to the fast radiative recombination and the surface reflection of the emitted radiation. As a result, in the virgin wafer the depletion region at the surface could be several μm thick, i.e. much larger than the additional thickness $t$. Therefore the changes in filtering of outgoing radiation are of no importance.

A quantitative estimate of $c$ and the nonradiative layer effect can be done by closer consideration of surface recombination effects. It has been shown [19] that these effects reduce the intensity of outgoing radiation by a factor $1/(1 + \tau \, S/l)$, describing an effective decrease of the luminescence intensity due to the loss of holes in the nonradiative layer. Here $\tau$ is the hole lifetime, $l$ is the effective thickness of the depletion region and $S$ is the surface recombination velocity. Though the exact values of these parameters is difficult to estimate, one can see that the increase of the surface recombination velocity above $S \cong 10^5$ cm/s can easily account for the nonradiative layer effect.

## Conclusions

We have studied thermal endurance of high-radiative-efficiency $n$-InP samples grown by the bulk and epitaxial HVPE technologies. The radiative efficiency of wafers grown by the traditional bulk Czochralski process catastrophically degrades under heating. This makes it impossible to use these wafers actively in optoelectronic devices that require subsequent overgrowth by epitaxial layers or other high-temperature processing. In contrast, HVPE grown layers demonstrate a very high radiative efficiency that does not degrade upon subsequent epitaxial overgrowth at a high temperature. We note that this finding is a very promising result for the future implementation of the "opaque" scintillator in InP material, as it permits utilization of its most remarkable property, the near perfect photon recycling.

## Acknowledgements

Part of this work was carried out with the support by the Domestic Nuclear Detection Office (DNDO) of the Department of Homeland Security, by the Defense Threat Reduction Agency (DTRA) through its basic research program, and by the New York State Office of Science, Technology and Academic Research (NYSTAR) through the Center for Advanced Sensor Technology (Sensor CAT) at Stony Brook.